\documentclass[12pt]{iopart}

\usepackage{graphicx}
\usepackage{epstopdf}
\usepackage{xcolor}
\begin{document}
\title[MW resonator for cold atom frequency standard ]{In-vacuum microwave resonator for a compact cold atom frequency standard}
\author{M. Givon$^1$, L. Habib$^1$, A. Waxman$^2$, Y. Bar-Haim$^1$, O. Amit$^1$ Y. Cina$^1$, G. Boneh$^1$, D. Groswasser$^1$, T. David$^3$, B. Levy$^2$, A. Stern$^2$ and R. Folman $^1$},
\address{$1.$ Ben-Gurion University of the Negev, 1 Ben-Gurion Blvd., Be'er-Sheva, 8410501 Israel}
\address{$2.$ AccuBeat Ltd., 5 Ha-Marpe St., Jerusalem, 9777405 Israel}
\address{$3.$ Quantum Art Ltd., 22 Einstein St., Nes Tziona, 7403686 Israel}
\ead{givonme@bgu.ac.il}
\begin{abstract}
A physics package for a compact cold atomic clock is hereby presented. The uniqueness of this package is its small dimensions that enable, for the first time, implementation of a primary cold atomic clock in a standard package of 3U height (=133\,mm). These dimensions are made possible by using an in-vacuum Microwave (MW) Loop Gap Resonator (LGR) whose length and diameter can be reduced from those of a typical resonator. Following our presentation of the design, we analyze the homogeneity of the MW field amplitude and phase. We find that the expected clock instability due to non-uniformity of the MW field is $\sim1.5\cdot10^{-14}$/day. Adding other clock errors, we estimate the total uncertainty  of a cold atomic clock  of this design will be around $\sim2\cdot10^{-14}$/day which results in a time drift of a few nanoseconds/day. Such a clock can serve as a primary-grade frequency reference, and may replace the cesium beam atomic clock and the GPS-disciplined rubidium atomic clocks.   
\end{abstract}
\ioptwocol
\section{Introduction}\label{sec:int} 
During the last fifteen years several groups have worked on the development of compact primary frequency standards based on cold atoms medium, usually referred to as ``cold atom clocks'' \cite{Quantum_phys}.
For example, in 2011 M\"{u}ller et al.~\cite{Brasil} developed an atomic clock based on a compact metal vacuum chamber simultaneously serving as a microwave (MW) cavity. In this clock cesium atoms are cooled using a magneto-optical trap (MOT) and then Ramsey interrogated by MW pulses while free-falling inside the chamber. They reached uncertainty of  $1.6\cdot10^{-14}$ for integration time of 1,000 seconds (with the central Ramsey fringe looked to an hydrogen maser).
 
 Muquans' Time and Frequency Division presented their ``MuClock''~\cite{Muclock}; a cold atom clock, based on isotropic cooling. They reached uncertainty of $1\cdot10^{-15}$ in less than 2 days integration time. SpectraDynamics Inc. matches this performance with its cold atom clock prototype~\cite{SD}, exhibiting uncertainty of $9\cdot10^{-16}$ in one day integration time. 

These clocks and several others~\cite{Elvin_I,Elvin_II,Lutwak,space,LGR_6_beam} demonstrate that cold atom clocks may suppress the uncertainty of the commercial cesium beam primary frequency standard (Model 5071A by Microchip) \cite{5071A,Fifty_years}, reaching uncertainty of $5\cdot10^{-15}$ in 5 days of integration. Having said that, the dimensions of the cold atom clocks presented so far are much larger than those of the 5071A, and therefore they cannot replace it for most applications, especially where portability and mobility are required. In particular, the height of these devices exceeds the 3U (=133\,mm) height of the Cs beam clock, which is a standard size for portable electronic equipment.

In this paper we provide a general description of the design of a novel compact physics core made to fit in a 3U box. This core is part of a cold atom clock prototype, and at the heart of this design is an in-vacuum MW resonator on which we focus here. Note that the design of the physics package must leave enough space (in a 19 inch rack, 3U slot) for all other components of the clock (lasers, optics and electronics). The structure of the rest of this paper is as follows: Section 2 details the design requirements of this physics core and the engineering solutions. In section 3 we focus on the detailed design of the MW resonator, while in section 4 we estimate the performance of an atomic clock based on this physics core. Section 5 contains the conclusion and outlook.      
\begin{figure}
  \includegraphics[width=\linewidth, clip] {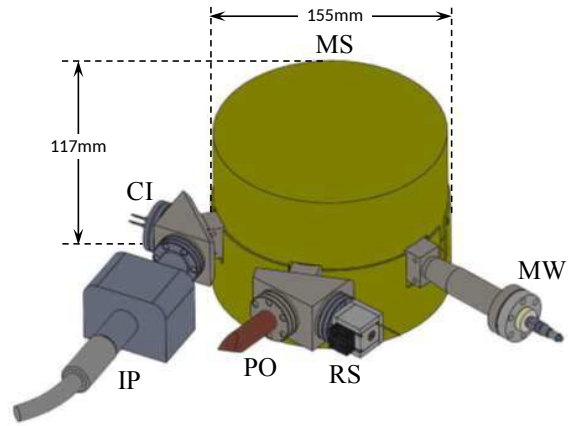}
  \caption{(Color online) Physics package - general view. MS - magnetic shielding (the vacuum chamber and the optical arrangement are inside the magnetic shielding); CI - feedthrough for current inputs; IP - ion pump; PO - pinch-off pipe (used for initial pumping); RS - rubidium source; MW - feedthrough for MW input. This device can fit in a standard  package of 3U height(=133\,mm).}
  \label{fig:full_sys}
\end{figure}


\section{System design}\label{sec:dis_con}

\subsection{Design constraints} 

The first phase in this atomic clock operation sequence is the cooling of $^{87}$Rb atoms in a pyramid MOT \cite{PYR_MOT} located inside an ultra-high vacuum (UHV) chamber. Then the cloud of cold atoms is Ramsey interrogated using MW irradiation during free-fall to accurately detect the frequency of the clock transition. The combination of atom cooling with MW interrogation in such a small volume is rather challenging, especially since these systems have contradictory requirements: while atom cooling demands a large window in the vacuum chamber for letting a wide laser beam pass through the atomic medium, the MW irradiation is best performing when transmitted to a resonator enclosed by metallic (reflecting) surfaces. Another conflict originates in the pyramid dielectric mirrors which are perfect for cooling atoms in a compact device, but absorb the MW irradiation and perturb the MW field homogeneity.

The spatial separation of the two tasks (cooling and MW interrogation) has been successfully demonstrated before~\cite{SD}; the atoms were cooled and then launched by laser light into a MW resonator located beneath the cooling zone. Naturally, the resulting physics package then exceeds the 3U size. We therefore decided to build a ``hybrid'' physics package and find other solutions to the conflicting requirements of the two functionalities, as elaborated below.    

\begin{figure}[h]
  \includegraphics[width=\linewidth] {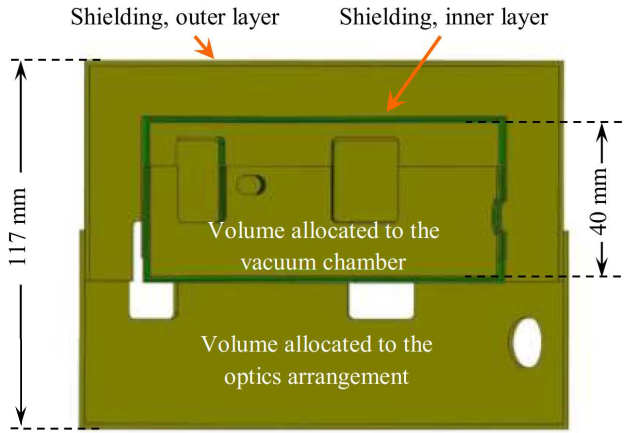}
  \caption{(Color online) Vertical cut showing volume allocation. The vertical cut through the two layers of magnetic shielding of the physics package of the atomic clock, showing volume allocation for the different sub-systems. The height of the vacuum chamber can not exceed 40mm.}
  \label{fig:cut_sys}
\end{figure}

After including the box's top and bottom covers, a height of $117\,\textrm{mm}$ remains for housing the vacuum chamber, the optics arrangement and the two-layer magnetic shielding. The optical set-up may be placed between the two shielding layers (Fig. \ref{fig:cut_sys}), leaving only $40\,\textrm{mm}$ free for the vacuum chamber itself.  

This rather small chamber must accommodate all the following elements: 
\begin{itemize}
  \item A MW resonator that is used to produce the two MW $\pi/2$ Rabi pulses of the Ramsey sequence. The typical inner dimension (this relates to both diameter and length) of such an element is usually around half a wavelength which is about $\rm22\,mm$ for $^{87}$Rb. 
  \item MOT coils that produce the magnetic gradients needed for spatial capturing the $^{87}$Rb atoms. The distance between these coils, which is required for the generation of a sufficiently high gradient necessary to capture the atoms, is  $\rm15-20\,mm$; adding the coils' thickness, the entire structure is about $\rm30\,mm$.
  \item C-field coils that produce a small constant and uniform magnetic field that is applied for removing the degeneracy of the clock transition, thereby leading to its spectral narrowing and improving the clock's stability. These coils surround the MOT coils and their length is about the same, so as they do not occupy any additional height. 
  \item An optical window for letting the laser light in, with special nonmagnetic sealing for UHV.  
  \item A pyramid structure generating retro-reflected beams for atom cooling. As mentioned above, locating the pyramid inside the resonator violates the field homogeneity, so we must place it above the resonator, lengthening our physics package by about $\rm6\,mm$.
\end{itemize}
 
 An extensive examination of several alternatives led us to the conclusion in that order to have a cold atom clock that fits a 3U ($=\rm133\,mm$ rack height, the MW resonator, the MOT and C-filed coils must all be inside the vacuum chamber. As a side benefit, this will decrease the power requirements of these coils.
   
Another set of limitations is related to the MW radiation. The magnetic component of the MW radiation must be aligned in the $z$ direction (along the symmetry axis of the chamber), and must be constant throughout the region occupied by the cold atoms. Any deviation from these requirements will degrade the stability of the atomic clock, as described in Section~\ref{Impact}.

Materials used to build the chamber and its inner components are also limited. They must be UHV compatible, have a relative magnetic permeability as close to 1 as possible, and apart from the coil wires, the MW resonator and the chamber body, have zero conductivity and low relative permittivity (dielectric constant).


\subsection{The MW resonator}\label{sec:MW}
\begin{figure}[h]
  \includegraphics[width=\linewidth, clip] {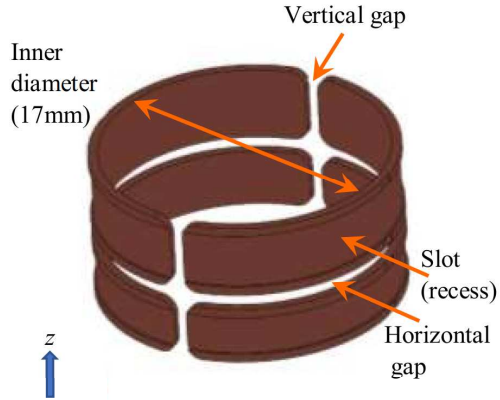}
  \caption{(Color online) The copper made Loop-Gap resonator (LGR). The resonance frequency and excited mode can be modified by changing several parameters: the inner diameter, the thickness and width of the copper half-rings, the depth of the slots and the width of the vertical and horizontal gaps. The plurality of parameters allows good optimization.}
  \label{fig:LGR_struc}
\end{figure}

The design that we chose for the MW resonator is known as the Loop-Gap Resonator (LGR) \cite{LGR1,LGR2,LGR3,LGR4}. This design has already been used for vapor-based atomic clocks \cite{T-Violetti,LGR-clock1,LGR-clock2}. A major advantage of this design is the ability to reduce the size of the resonator below the ordinary limit of $\lambda$/2. Another advantage is the fact that there are several parameters (see Fig. \ref{fig:LGR_struc}) that one can use to modify the properties of the MW field. In addition, the resonator is open at both ends, enabling the laser light free passage.
  
\begin{figure}[h]
  \includegraphics[width=\linewidth, clip] {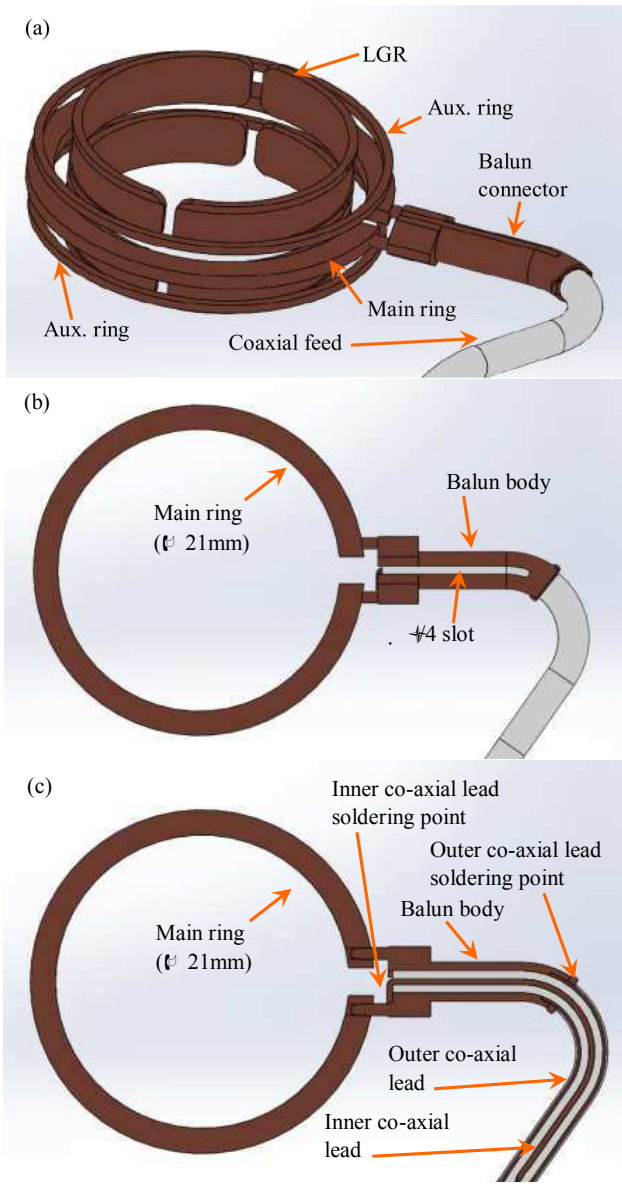}
  \caption{(Color online) In-vacuum excitation of the LGR. a. Shown are the main excitation ring fed by the ``balun'' (balanced to unbalanced) connector and two auxiliary (parasitic) rings. b. Top view of the main ring and the balun body, showing the $\lambda/4$ long slot (a similar slot is on the other side of the balun body). c. Section view of the main ring, the balun body and the coaxial feed. Note that the outer coaxial lead is soldered (with special UHV compatible soldering materials) to the balun body on its far end while the inner lead is soldered to its near end. This design improves the axial symmetry of the MW radiation.}
  \label{fig:Excitation}
\end{figure}

The MW input feeds a set of excitation rings through a specially designed ``balun'' (balanced to unbalanced) connector (see Fig. \ref{fig:Excitation}). 

The purpose of the balun is to force the MW current to flow in the main excitation ring with equal intensity in both directions (clockwise and counterclockwise), thus ensuring radial symmetry of the magnetic component of the MW field induced by the LGR. The purpose of the two auxiliary rings is to improve the field's axial uniformity. 


\subsection{The inner structure and the chamber}\label{sec:Inner}
\begin{figure}[h]
  \includegraphics[width=\linewidth, clip] {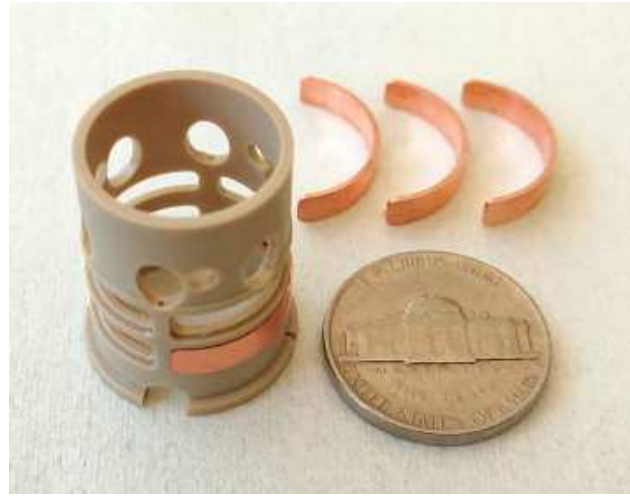}
  \caption{(Color online) The LGR components and holder. The holder is shown with windows that fit the LGR components. One of the LRG components is inserted in its window. The nickel coin is for size reference.}
  \label{fig:INN_struc}
\end{figure}

The purpose of the inner structure is to hold in position all the components of the LGR, the excitation rings, the MOT and C-field coils. Fig. \ref{fig:INN_struc} shows the central component of the inner structure - the LGR holder with its four LGR components, each to be inserted in the windows cut into the LGR holder. The LGR holder is made of PEEK (polyether ether ketone), a UHV-compatible, high-temperature thermoplastic polymer with low relative permittivity ($\sim$3.3) and relative magnetic permeability close to 1. 

\begin{figure*}[t!]
  \includegraphics[width=\linewidth, clip] {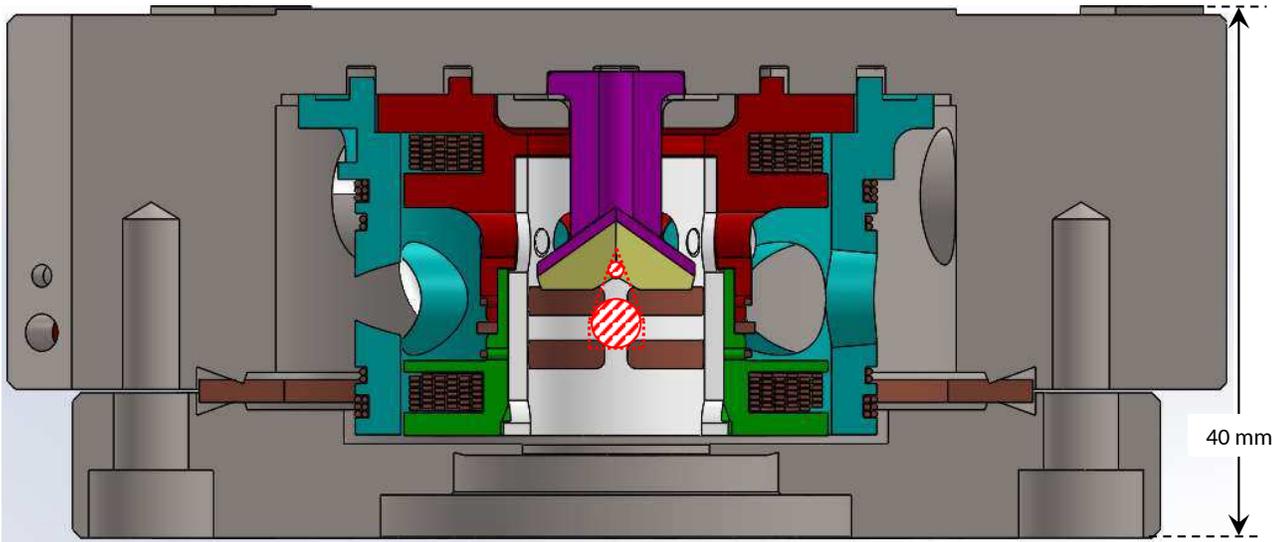}
  \caption{(Color online) Color-coded section through the clock's chamber and inner structure. Grey - the titanium body of the chamber. Note the opening at the bottom of the chamber for the laser beams. Purple - the glass column that holds the MOT mirrors of the inverted pyramid; purple covered with gold color - the dielectric glass mirrors; reddish-brown - the copper LGR, excitation rings and coils; white - a PEEK component that holds the LGR; red and green - PEEK components that hold the MOT coils and the excitation rings; azure - a PEEK component that holds the C-field coils. The small red striped circle near the center of the mirrors indicates the location where the cold-atom cloud is created, and the larger red striped circle under it shows the location of the cloud after a free-fall of $~$30 milliseconds. The dashed red lines mark the area where the intensity of the MW field should be constant and the direction of the magnetic components of the MW field should be vertical.}
  \label{fig:cut_full_sys}
\end{figure*} 

Fig.~\ref{fig:cut_full_sys} shows the components of the inner structure and their relation to the vacuum chamber and to the pyramid MOT mirrors.  All these components affect the MW field created by the LGR.
Naturally, the effect of the metallic components is dominant, but one cannot ignore the effect of the PEEK and glass components near the LGR.

\section{LGR design and MW simulation}\label{sec:MW_sim}
The LGR design has several purposes. One is to create, in the area limited by the dashed red lines in Fig. \ref{fig:cut_full_sys}, a MW field with constant intensity of its magnetic component. A second purpose is that the magnetic component of the MW field will be aligned vertically. In addition, the LGR must have a resonance at 6.834\,GHz (the frequency of the clock transition in $^{87}$Rb), so that most of the power fed to the LGR will be transmitted to the atoms and not reflected back. For robustness, we prefer a wide resonance (low Q), so that minor changes in the components of the inner structure will not cause large changes in the intensity of the MW field at the location of the atoms.         

\subsection{MW simulation and MW measurements}\label{sec:MW_sim}
\begin{figure}
  \includegraphics[width=\linewidth, clip] {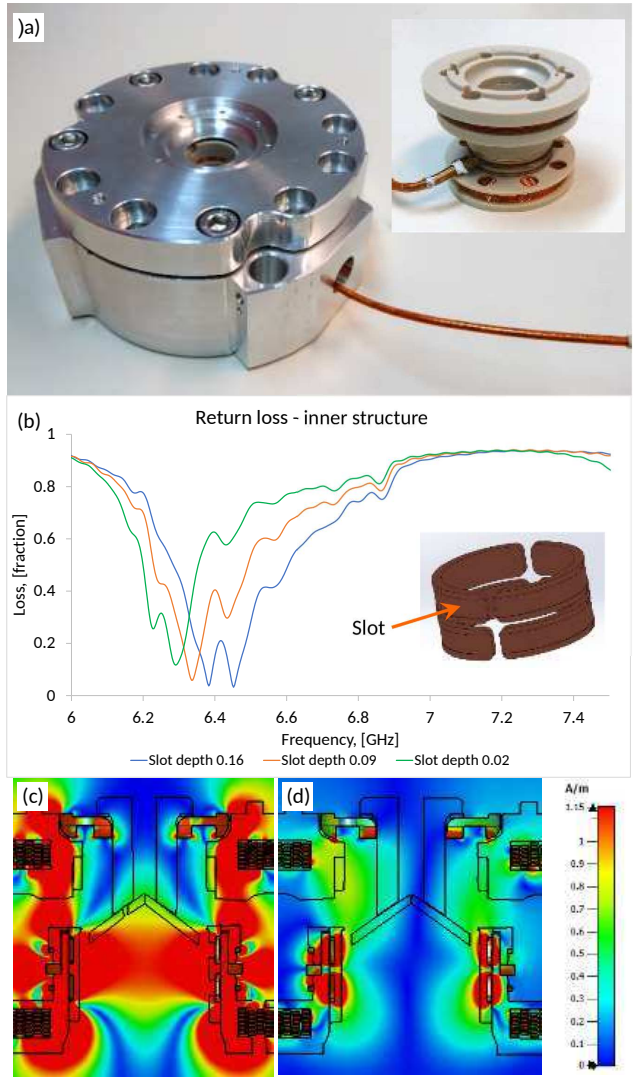}
  \caption{(Color online) MW simulation. a. The mock-up chamber built to partly verify the MW simulation results. Insert: the inner structure. b. Plots of the return loss reflected from the inner structure, simulated in free space (stand-alone inner structure). The simulation is done for 3 values of the depth of the slot in the LGR. Inset: the LGR, showing the slot. Note that a dip in the return loss indicates a peak in the MW energy transmitted to the volume occupied by the atoms. c. Color map of the simulated amplitude of $H_z$ (the magnetic component of the MW field parallel to the symmetry axis; see text). d. Color map of the simulated amplitude of $H_y$ (the component perpendicular to symmetry axis; see text).}
  \label{fig:sim_results}
\end{figure}

The mechanical and MW designs were developed in parallel. Initially, we made some calculations of a simple LGR structure to provide a first estimation of the resonance frequency and the mode structure of the resonator, following which we developed an iterative process. First, we prepare a model of the mechanical design using SolidWorks~\cite{Solid}. Next, we import the model to the MW simulation program CST \cite{CST} and simulate the MW field inside the chamber. In particular, we simulate the return loss reflected by the inner structure (see Fig.~\ref{fig:sim_results}a) and the amplitude of the magnetic component of the MW field in the vicinity of the cold atom cloud (inside the dashed red lines in Fig.~\ref{fig:cut_full_sys}). We then study the results and modify the mechanical design and/or the LGR components as needed to get the MW simulation results closer to our requirements. Naturally, we sometimes had to modify the mechanical design for other reasons. For every such change, we ran a MW simulation to verify that the change did not adversely affect the MW field. Based on the combined results of the mechanical design and MW simulations, we produced all the components (including a ``mock-up'' chamber) and assembled the chamber with the mirrors' pyramid and all the components of the inner structure. We will refer to this design as Rev1 (Fig.~\ref{fig:sim_results}a).

In Fig. \ref{fig:sim_results}b we present simulated results of the return loss reflected by the inner structure in three different configurations. Each configuration has LGR components with different slot depth, and we see that by changing the slot depth we can control the location of the dip in the return loss by $\sim \pm 100$\,MHz. Figs.~\ref{fig:sim_results}c and \ref{fig:sim_results}d show the simulated amplitude (in A/m) of the magnetic component of the MW field parallel (perpendicular) to the symmetry axis, respectively, at a MW frequency of $6.375$\,GHz. We see that the mode the LGR excites at this frequency is the mode we need, with most of the magnetic component aligned parallel to the chamber's symmetry axis ($z$ axis in Fig. \ref{fig:LGR_struc}). However, the simulated resonance frequency is $\sim450$\,MHz lower than our target, $6.834$\,GHz.          

\begin{figure}[h]
  \includegraphics[width=\linewidth, clip] {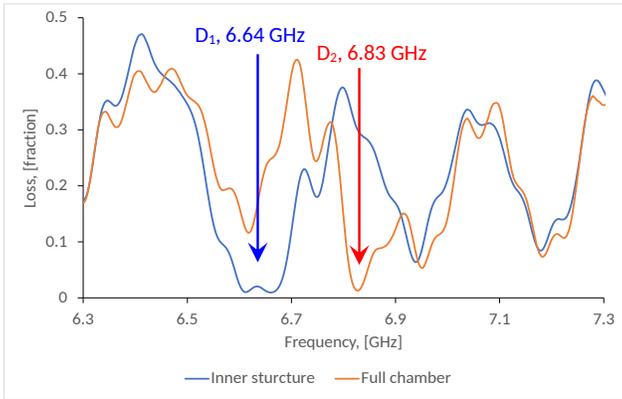}
  \caption{(Color online) Measured return loss. Blue: return loss reflected by the inner structure. The resonance dip marked $D_1$ is the only dip that reacts to the interference test (see text). Red: return loss reflected by the full chamber. The resonance dip marked $D_2$ is the only dip that reacts to the interference test (see text). Thus we conclude that at the frequencies of the $D_1$ and $D_2$ dips the LGR excite the MW radiation modes we need in the inner structure and the full chamber, respectfully. (See Fig.~\ref{fig:sim_results} and text).}
  \label{fig:S11_full}
\end{figure}

We measured the return loss reflected by the inner structure as well as by the fully assembled chamber  using a network analyzer (Rohde $\&$ Schwarz ZND vector network analyzer). We present the results in Fig.~\ref{fig:S11_full}.      

We see several dips in the measured return loss. We use an interference test to identify which dip is related to the MW mode we need: we insert a material with high relative permittivity into the area below the mirrors' pyramid (via the opening in the chamber, see Figs.~\ref{fig:cut_full_sys} and~\ref{fig:LGR_struc}a), and monitor the response of the network analyzer to that interference. We note that most of the dips do not react at all to the interference test, while the the dips marked ${D_1, D_2}$ in Fig. \ref{fig:S11_full} almost disappear under that test. We therefore conclude that at the resonance frequencies of these dips the LGR excite the MW radiation mode we need (see Fig.~\ref{fig:sim_results}).       

\subsection{Tuning the resonance frequency}\label{sec:updated}
The results of many simulations and tests have led us to the following conclusions:
\begin{itemize}
\item The relevant resonance frequency of the full chamber ($D_2$ in Fig \ref{fig:S11_full}) is higher by $\sim 200\,\textrm{MHz}$ than the relevant resonance frequency of the inner stricture ($D_1$ in Fig \ref{fig:S11_full}).
\item The \textbf{measured} relevant resonance frequency of the full chamber ($D_2$ in Fig \ref{fig:S11_full}) is higher by $\sim 450\,\textrm{MHz}$ than the \textbf{simulated} one.
\end{itemize}
These conclusions served as the basis for the design. Following several MW simulations of the system wherein the dimensions of several parameters were scanned (the diameter and location of the LGR, the vertical and horizontal gaps and the slot depth [see Fig. \ref{fig:LGR_struc}]), we reached a robust design whereby the simulated resonance is at $6.4\,\textrm{GHz}$. We therefore anticipate the measured resonance will be near the required $6.834\,\textrm{GHz}$. To fine-tune the resonance frequency, we prepared several LGR components with some minute changes in their dimensions (e.g. the slot depth), perform MW measurements of the location of the resonance during the assembly process and, if needed, replace the LGR components until we reach the required resonance frequency.               

\section{Impact of the MW field on the clock performance}\label{Impact}

Below we analyze the effect of the field amplitude and phase inhomogeneity on the clock performance. 

As illustrated by the two circles in Fig. \ref{fig:cut_full_sys}, the atoms fall a few millimeters during the Ramsey sequence ($\rm2\,mm$ for $\rm20\,ms$ dark time). Since the MW magnetic field varies in space, as shown in Fig.~\ref{fig:sim_results}c-d, the atoms experience a different MW field -- and therefore a different Rabi frequency -- at the second Ramsey pulse compared to the first. Given that the first pulse area (which is the Rabi frequency times the pulse time) is exactly $\pi/2$, this necessarily means the second one isn't. As a result, the contrast of the Ramsey fringe is reduced, thereby degrading the clock stability.

\begin{figure*}[t!]
  \includegraphics[width=\linewidth, trim= 0 3.3cm 0 2.5cm, clip] {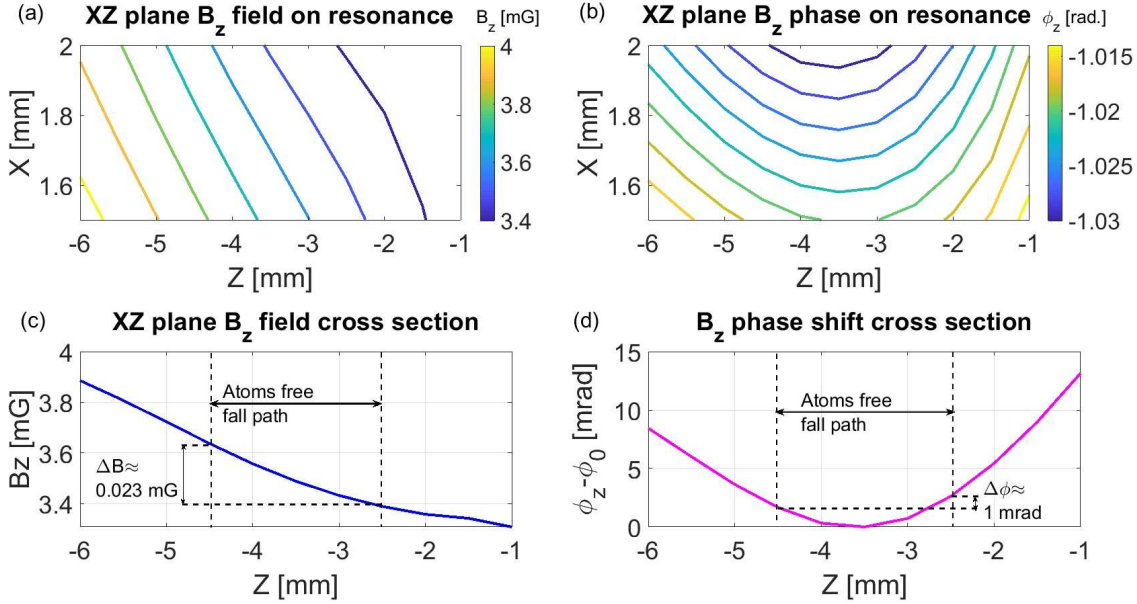}
  \caption{(Color online) Phase and amplitude distribution along the free-fall trajectory of atoms during the Ramsey sequence. In (a) and (b) a plane stretching from the pyramid symmetry axis to the edge of the atomic cloud is shown, while in (c) and (d) a cross section going through the centers of the initial and final cloud is analyzed. The range of the atoms' trajectory is marked on these graphs.}
  \label{fig:paper_fig_XZ}
\end{figure*}

However, this effect is negligible in our case: we have found that the field changes by only $\sim6\%$ along the falling path of the atoms (see Fig. \ref{fig:paper_fig_XZ}), which means the contrast, as well as the stability of the clock, is reduced by less than 1$\%$. 

In addition, we can experimentally evaluate the field variation by measuring the Rabi frequency of the atoms as a function of their time of flight. Then, we can adjust the duration of the second pulse so that its area is $\pi/2$ as well.  
However, phase inhomogeneity is more devastating to the clock performance than amplitude non-uniformity: while the latter attenuates the Ramsey fringe amplitude, the former shifts its central frequency, leading to a clock frequency offset.   

\begin{table*}[t]
    \centering
   \begin{tabular}{p{0.3cm}p{4.8cm}p{3.6cm}p{3cm}p{3cm}}
$\#$ & Noise source [ref]  & Relevant Quantity & Frequency shift & Uncertainty \\ 
\hline
1 & MW phase inhomogeniety & phase shift ($\Delta\phi$) & 1.0E-12 & 1.5E-14   \\
2 & MW switching \cite{5116858} &  phase shift ($\Delta\phi$)  & 5.0E-15 & 5.0E-16  \\
3 & 2$^{\textrm{nd}}$ order Zeeman \cite{5507655} & Magnetic field (B$_0$) & 6.0E-10 & 1.0E-14  \\
4 & Detection noise \cite{8920050} & laser power ($P_l$) & N.A & 1.0E-14   \\
5 & Off res. light shift \cite{5507655} & laser frequency ($\nu_l$) & N.A & 1.0E-15  \\
\hline
  &Total&  & 6.0E-10 &2.1E-14 \\ 
\end{tabular}
    \caption{An estimated noise budget for our compact atomic clock. Note: the uncertainty is for $\tau<\textrm{10}^\textrm{5}$\,s. The frequency shift and uncertainty due to MW inhomogeniety is derived in this section. The rest of the noise sources are analyzed in the references attached. The shifts and uncertainties of the relevant quantities are either measured (rows 3 and 4 - unpublished results), simulated (row 1), or based on typical results from literature (rows 2 \cite{5116858} and 5 \cite{5507655}).}
    \label{tab:uncertainties}
\end{table*}

In the Bloch sphere representation, a MW phase variation between the first and second pulses leads to a change of the Bloch vector's rotation axis (in the $xy$-plane) during the second Ramsey pulse. Such a change is equivalent to a rotation, during the Ramsey interrogation, of the Bloch vector around the~$z$ axis when pointing to the equator. However, such a rotation also occurs when the MW radiation frequency is detuned from the atomic transition frequency. Moreover, this rotation enables us to evaluate the MW frequency error relative to the atomic transition frequency, since it grows as $\delta T_r$, with $\delta$ being the detuning and $T_{r}$ the Ramsey interrogation time. In other words, we cannot distinguish between a clock frequency error and a MW phase shift between the two Ramsey pulses, as shown by   
\begin{equation}
\label{eq:delta}
    \delta_{m}T_r =  \delta_{r}T_r+\Delta\phi_{MW},
\end{equation}
where $\delta_{m}$ and $\delta_{r}$ are the measured and actual detunings, respectively, and $\Delta\phi_{MW}$ is the MW phase shift. 
Now we can extract the clock's relative frequency shift induced by the MW phase shift:
\begin{equation}
\label{eq:ferror}
  y_{ps}= \frac{\delta_{m}-\delta_{r}}{\nu_0}=\frac{\Delta\phi_{MW}}{2\pi\nu_0T_r},
\end{equation}
where $\nu_0=6.834\,\textrm{GHz}$ is the atomic transition frequency.

Accounting for a phase shift of 1\,mrad (see Fig. \ref{fig:paper_fig_XZ}) between the initial point (first $\pi$/2 pulse) and the final point (second $\pi$/2 pulse), we have calculated the relative frequency shift to be $\sim1\cdot10^{-12}$ where we have considered a Ramsey time of $\rm20\,ms$. We can easily cancel this shift during the initial calibration of the clock.

Nevertheless, variations in the initial position of the cloud due to changes in the magnetic trapping potential or cooling light intensity fluctuations add uncertainty to this shift. To estimate it we first derive the phase shift gradient from the simulation data: ${d\phi}/{dr}\approx3\,\mu\textrm{rad}/\mu\textrm{m}$. Next, we fit several hundreds of atomic absorption images taken during a day to a 2D Gaussian waveform, and extract the center of mass location. The RMS deviation of the cloud location during a day is found to be $\sim 7\,\mu\textrm{m}$, meaning the RMS phase deviation is $\sim 21\,\mu\textrm{rad}$. According to Eq.[\ref{eq:ferror}] the clock frequency shift due to MW phase inhomogeneity is then $\sim1.5\cdot10^{-14}$ per day.

In Table \ref{tab:uncertainties} we list the dominant mid and long term noises effecting our clock. The shifts and uncertainties are based on measurements or calculations of the relevant physical quantities (e.g. magnetic field). The relation between each quantity and the clock error is given in the reference appears by the noise source. The total uncertainty we predict
is about $2.1\cdot10^{-14}$ per day.  

\section{Conclusion and outlook}\label{conc}
The aim of our work was to design and build a compact physics package for cold atom clock. For commercial reasons, it is important that the clock should fit a 3U slot of a standard 19 inch rack (height $133\,$mm). Our main achievement, fitting the pyramid mirrors, the MOT coils, the MW resonator and the C-field coils in a 40\,mm high structure (see Fig.~\ref{fig:cut_full_sys}) made it possible to meet this goal. The additional components (the optical arrangement and the double-layer magnetic shielding) enlarge the package to a height of just 117\,mm (see Fig.~\ref{fig:full_sys}). To the best of our knowledge, the height of the presented design is at least three time smaller then any other design of a cold atom clock.

This paper focuses on a detailed analysis of the in-vacuum MW resonator, and its influence on the clock's stability. Despite of its small size, our estimates show that a clock based on the presented physics core will have an uncertainty of $\sim2\cdot10^{-14}$ per day, which may go down further with longer integration time. The resulting time drift of a cold-atom clock of this design will be in the range of a few nanosecond/day. Such a clock can serve as a primary frequency reference, and replace for several applications the cesium beam atomic clock and GPS-disciplined rubidium atomic clocks.

We have already built and operated two lab-scale prototypes testing several aspects of this design (unpublished), and we are now in the final stages of constructing an engineering prototype that will enable the validation of the clock's accuracy and stability. 

We note here that orders of magnitude better accuracy and stability will eventually require moving to an optical frequency standard, but this requires completely new technologies which are very different and much more complex, and are currently very bulky and expensive. Thus we believe that the cold atom microwave clock will still be the work horse of the time keeping industry for the foreseeable future.

\section*{Acknowledgement}\label{ack}
This paper is based upon research supported in part by the U. S. Office of Naval Research under award number N62909-18-1-2159, in part by the Israel Innovation Authority under award number 74482 and in part by the Israel Science Foundation, grant number 3470/21.   
\section*{References}
\providecommand{\newblock}{}

\end{document}